\documentclass[useAMS,usenatbib]{emulateapj}
\usepackage{lineno}
\usepackage{graphicx}
\usepackage{amssymb}
\usepackage{natbib}
\usepackage{draftwatermark}
\usepackage{comment}

\shorttitle{Lag behavior in GRS 1915+105}
\shortauthors{Pahari et al.}

\begin{document}

\SetWatermarkAngle{45.0}
\SetWatermarkLightness{0.9}
\SetWatermarkFontSize{10.0 cm}
\SetWatermarkText{}

\title{Comparison of time/phase lags in the hard state and plateau state of GRS 1915+105}
\author{Mayukh Pahari\altaffilmark{1}, Joseph Neilsen\altaffilmark{2,3}, J S Yadav\altaffilmark{1}, Ranjeev Misra\altaffilmark{4}, Phil Uttley\altaffilmark{5}} 
\altaffiltext{1}{Tata Institute of Fundamental Research, Homi Bhabha Road, Mumbai, India; \texttt{mp@tifr.res.in(MP)}}
\altaffiltext{2}{Einstein Fellow, Boston University, Boston, MA 02215, USA}
\altaffiltext{3}{Massachusetts Institute of Technology, Kavli Institute for Astrophysics, Cambridge, MA 02139, USA}
\altaffiltext{4}{Inter University Center for Astronomy and Astrophysics, Pune, India}
\altaffiltext{5}{Astronomical Institute, ``Anton Pannekoek,'' University of Amsterdam, Science Park 904, 1098-XH Amsterdam, The Netherlands}

\linenumbers

\begin{abstract}
We investigate the complex behavior of energy- and frequency-dependent
time/phase lags in the plateau state and the radio-quiet hard ($\chi$)
state of GRS 1915+105. In our timing analysis, we find that when the
source is faint in the radio, QPOs are observed above 2 Hz and
typically exhibit soft lags (soft photons lag hard photons), whereas
QPOs in the radio-bright plateau state are found below 2.2 Hz and
consistently show hard lags. The phase lag at the QPO frequency is
strongly anti-correlated with the QPO frequency, changing sign at 2.2
Hz. However, the phase lag at the frequency of the first harmonic is
positive and nearly independent of frequency at at $\sim0.172$ rad,
regardless of the radio emission. The lag-energy dependence at the
first harmonic is also independent of radio flux. However, the lags at
the QPO frequency are negative at all energies during the radio-quiet
state, but lags at the QPO frequency during the plateau state are positive at all energies and show a 
`reflection-type' evolution of the lag-energy spectra with respect to the radio-quiet state. 
The lag-energy dependence is roughly logarithmic, but there is some evidence for a break around $4-6$ keV. 
Finally, the Fourier frequency-dependent phase lag spectra are fairly flat during the plateau state, 
but increase from negative to positive during the radio-quiet state. 
We discuss the implications of our results in the light of some generic models.

 \end{abstract}  


\section{Introduction}

More than 20 years have passed since the discovery of the Galactic black hole X-ray binary GRS 1915+105, yet this well-studied black hole continues to reveal new physical insights into the phenomenon of black hole accretion. Apart from its extraordinary activities (like superluminal motion of jets) in the radio and infrared, it exhibits diverse X-ray variability patterns (e.g., \citealt{b81,Belloni00,b32,H05}), in their light curves and color-color diagrams. Each of these variability pattern can be decomposed into three basic states: a low luminosity, spectrally hard state (dominated by non-thermal emission) with a disappeared inner accretion disk (state C), a high luminosity, spectrally soft state (dominated by the disk blackbody) with an inner disk extending to the innermost stable circular orbit (state B), and a low luminosity, spectrally soft state with spectrum similar to the high-soft state (state A). There is no clear relationship between these A, B, and C states and the canonical states of black hole binaries (e.g.\ \citealt{b25}), although \citet{b34} noted that all three of these states show some similarities to the Very High State.

\begin{figure}
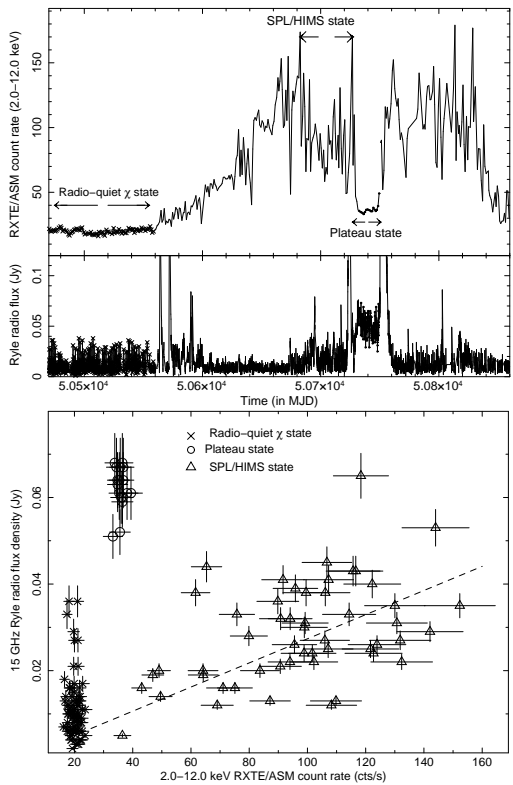

\centerline{\includegraphics[width=0.295\textwidth,angle=-90]{f1a}}
\centerline{\includegraphics[width=0.275\textwidth,angle=-90]{f1b}}
\caption{Left: One-day average $2-12$ keV X-ray lightcurve of GRS 1915+105 from the \textit{RXTE} All-Sky Monitor (ASM, top) and 15 GHz radio flux from the Ryle telescope. Right: the relation between ASM count rate and radio flux density. In both panels, radio-quiet $\chi$ states are shown as crosses, plateau states are shown as empty circles, and steep power law/hard intermediate states (SPL/HIMS) are shown as triangles.\label{fig:rx}}
\end{figure}

Among different variability classes, the closest analog to the canonical black hole ``low hard state" (LHS) is the $\chi$ class, which is found exclusively in state C \citep{Belloni00}. This analogy is somewhat complicated by the diversity of spectrally hard states in GRS 1915+105. Pooley \& Fender (1997) and \citet{b29} discovered that while some spectrally hard states are relatively radio faint, other states, which they called ``plateau'' states, are associated with bright, steady radio emission at 15 GHz (see also \citealt{b26}; Harmon 1997 for earlier discussions of plateaus). As described by \citet{b25} and references therein, the plateau state has historically arisen on time scales of days, appearing after sharp decreases and increases in the X-ray flux and radio flux density, respectively. A typical value of the plateau-state radio flux density is $\sim50-100$ mJy (\citealt{b29}; Figure \ref{fig:rx}).

The diversity of these spectral states is also reflected in their X-ray emission: the $\chi$ state itself has subclasses, called $\chi_1-\chi_4,$ which were first identified as distinct, long instances of the C state (\citealt{Belloni00}). In this classification, the $\chi_1$ and $\chi_3$ states, which are typically brighter and softer than $\chi_2$ and $\chi_4,$ happened to coincide with the radio plateau states. For this reason, we will refer to the radio-bright, X-ray bright $\chi$ states as plateau states, and their radio-faint analogs as the ``radio-quiet $\chi$ state'' for the rest of the paper.  Again, neither of these states is the same as the canonical black hole LHS. For example, as detailed by \citet{b25}, plateau states have a characteristic luminosity comparable to the Eddington limit ($L_{\rm Edd}$) and steeper spectra ($\Gamma\sim1.8-2.5$; \citealt{b28} and references therein), relative to the canonical LHS (with $\Gamma\sim 1.5$-$1.7$ and $L\lesssim 0.1L_{\rm Edd}).$ 

\begin{figure}
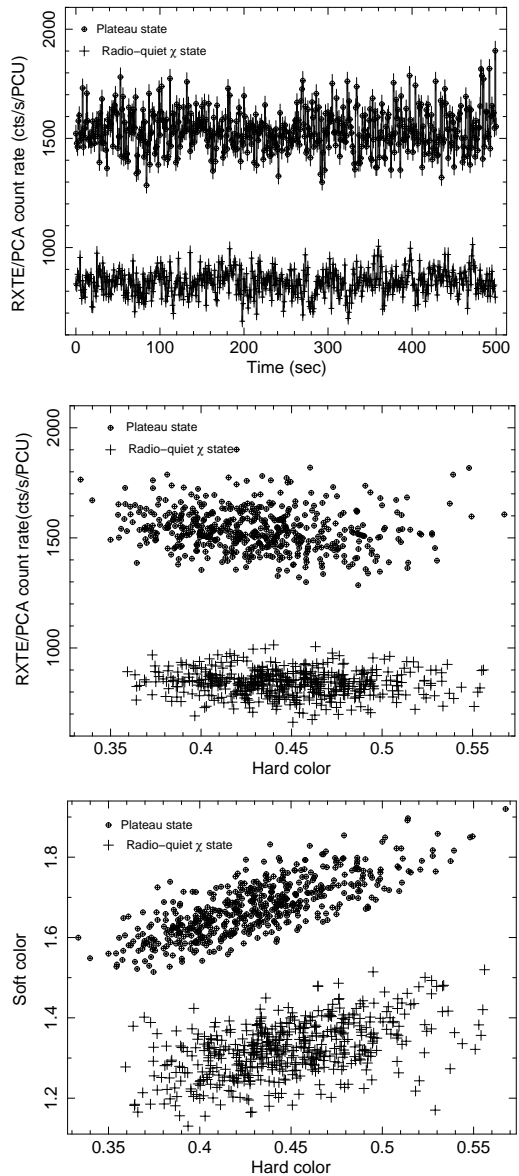

\centerline{\includegraphics[width=0.27531\textwidth,angle=-90]{f2a}}\vspace{3mm}
\centerline{\includegraphics[width=0.27531\textwidth,angle=-90]{f2b}}\vspace{3mm}
\centerline{\includegraphics[width=0.27531\textwidth,angle=-90]{f2c}}
\caption{Top: \textit{RXTE}/PCA $2-60$ keV light curves of example plateau and radio-quiet $\chi$ states. Middle: hardness-intensity diagram. Bottom: color-color diagram. Symbols are as in Figure \ref{fig:rx}. Error bars are $<6.5\%$ of the actual data values.\label{fig:lchr}}
\end{figure}

In timing studies, the $\chi$ state is known to exhibit strong broadband noise and $0.1-10$ Hz QPOs, which exhibit particularly complex time and phase dependence. For example, \citet{Reig00} found a strong anti-correlation between the QPO frequency and phase lags, such that the lag actually switches from positive to negative for QPOs around 2 Hz (see also \citealt{b20}). Because the negative lags were also associated with the highest count rates and the softest spectra, the authors argued that the timing behavior could possibly be explained by the presence of an optically thick scattering medium and a geometrically thick disk. \citet{Muno01} demonstrated that the QPO frequency decreases and continuum phase lags increase (from negative to positive) with increasing radio flux; the radio optical depth also appears to play a role in these relationships. However, it is not clear that a jet origin for QPOs is indicated, since the QPO amplitude is not sensitive to the radio flux (\citealt{Yan13}). In any case, the complex timing behavior presents a challenge to models that predict both X-ray variability and X-ray spectra (\citealt{Muno01}).

Thus, the relationship between the plateau state and the radio-quiet $\chi$ state in GRS 1915+105 remains of interest. How do the relevant accretion processes (and the associated spectral and timing properties) change with X-ray and radio luminosity? In this work, we explore the connections between these states through timing and spectra analysis of six radio-quiet $\chi$ states and six plateau states observed by the \textit{Rossi X-ray Timing Explorer (RXTE)}. We focus on the energy- and frequency-dependent lags, since these seem most likely to provide fruitful insights into the behavior of the accretion flow (owing to the radial dependence of both variability time scales and the X-ray spectrum). We find that the lags in radio-quiet $\chi$ and plateau states have very different energy and frequency dependence, although the phase lag at the first harmonic appears to be independent of state. We discuss our timing results in the context of prior results in the literature.

\begin{deluxetable*}{cccccccccccc}
\tabletypesize{\scriptsize}
\tablecaption{Observations of GRS 1915+105\label{tbl:dist}}
\tablewidth{0pt}
\tablehead{
\colhead{}  &
\colhead{}  & 
\colhead{}  &
\colhead{$\nu_0$}  &
\colhead{QPO$_0$}  &
\colhead{}  &
\colhead{$\nu_1$}  &
\colhead{QPO$_1$}  &
\colhead{}  &
\colhead{$S_{\rm 15 GHz}$} &
\colhead{$\alpha$} \\
\colhead{Obs}  &
\colhead{ObsID}  & 
\colhead{MJD}  &
\colhead{(Hz)}  &
\colhead{RMS (\%)}  &
\colhead{$Q_0$}  &
\colhead{(Hz)}  &
\colhead{RMS (\%)}  &
\colhead{$Q_1$}  &
\colhead{(mJy)} &
\colhead{(10$^{-2}$)}
}

\startdata

h1  & K-07-00 &  50436.78 & 3.09 $\pm$ 0.04 & 10.4 $\pm$ 0.7 & 6.31 $\pm$ 0.08 & 6.24 $\pm$ 0.06 & 4.51 $\pm$ 0.05 & 4.40 $\pm$ 0.09 & 9.5 $\pm$ 0.9 & -2.3 $\pm$ 0.4 \\ 
h2  & K-10-00 & 50456.32 & 2.79 $\pm$ 0.03 & 10.2 $\pm$ 0.4 & 7.97 $\pm$ 0.09 & 5.64 $\pm$ 0.08 & 5.62 $\pm$ 0.09 & 5.64 $\pm$ 0.07 & 5.8 $\pm$ 0.4 & -1.4 $\pm$ 0.2\tablenotemark{a}  \\ 
h3  & K-20-00 & 50525.04 & 3.22 $\pm$ 0.05 & 9.46 $\pm$ 0.09 & 7.13 $\pm$ 0.09 & 6.42 $\pm$ 0.04 & 5.4 $\pm$ 0.2 & 4.5 $\pm$ 0.2 & 4.6 $\pm$ 0.4 & -1.6 $\pm$ 0.1\tablenotemark{a}\\ 
h4  & T-19-00 & 53214.95 & 2.09 $\pm$ 0.04 & 13.8 $\pm$ 0.6 & 13.9 $\pm$ 0.4 & 4.14 $\pm$ 0.03 & 4.79 $\pm$ 0.02 & 15.9 $\pm$ 0.8 & 14 $\pm$ 3 & -2.3 $\pm$ 0.2 \\ 
h5  & O-35-00 & 51081.23 & 2.43 $\pm$ 0.04 & 12.0 $\pm$ 0.6 & 6.57 $\pm$ 0.08 & 4.88 $\pm$ 0.08 & 6.89 $\pm$ 0.07 & 4.1 $\pm$ 0.1 & 7.4 $\pm$ 0.5 & -2.1 $\pm$ 0.1 \\ 
h6  & Q-09-00 & 51366.32 & 2.13 $\pm$ 0.06 & 14.2 $\pm$ 0.4 & 9.2 $\pm$ 0.1 & 4.21 $\pm$ 0.03 & 4.58 $\pm$ 0.02 & 11.7 $\pm$ 0.3 & 16 $\pm$ 3 & -2.2 $\pm$ 0.1\\  
\hline \\ [-1.5ex]
\vspace{1mm}p1 & I-25-00  & 50283.49 & 1.11 $\pm$ 0.03 & 11.7 $\pm$ 0.3 & 8.5 $\pm$ 0.1 & 2.19 $\pm$ 0.03 & 2.34 $\pm$ 0.04 & 18 $\pm$ 1 & 106 $\pm$ 3 & 13.1 $\pm$ 0.7  \\ 
p2 & K-50-00  & 50735.55 & 0.83 $\pm$ 0.03 & 10.7 $\pm$ 0.3 & 8.4 $\pm$ 0.3 & 1.65 $\pm$ 0.04 & 3.83 $\pm$ 0.06 & 15.1 $\pm$ 0.9 & 81 $\pm$ 5 & 3.4 $\pm$ 0.4 \\          
p3 & S-50-00  & 52572.19 & 0.89 $\pm$ 0.04 & 11.3 $\pm$ 0.4 & 8.7 $\pm$ 0.1 & 1.76 $\pm$ 0.04 & 2.51 $\pm$ 0.03 & 17.5 $\pm$ 0.6 & 158 $\pm$ 7 & 12 $\pm$ 1 \\ 
p4 & S-52-00  & 52587.36 & 1.39 $\pm$ 0.02 & 11.9 $\pm$ 0.3 & 9.86 $\pm$ 0.02 & 2.74 $\pm$ 0.02 & 2.82 $\pm$ 0.02 & 13.7 $\pm$ 0.5 & 112 $\pm$ 7 & 3.5 $\pm$ 0.4 \\ 
p5 & N-10-00  & 54567.31 & 1.59 $\pm$ 0.03 & 12.4 $\pm$ 0.8 & 8.8 $\pm$ 0.6 & 3.14 $\pm$ 0.06 & 1.52 $\pm$ 0.07 & 18.4 $\pm$ 0.8  & \nodata & 10 $\pm$ 1\tablenotemark{a} \\ 
p6 & O-20-00  & 50957.35 & 0.71 $\pm$ 0.02 & 9.5 $\pm$ 0.4 & 8.6 $\pm$ 0.3 & 1.39 $\pm$ 0.03 & 3.02 $\pm$ 0.08 & 19 $\pm$ 1 & \nodata & 9 $\pm$ 2 \\ [-1.5 ex]
\enddata
\tablecomments{Observation details of radio-quiet $\chi$ state (h1 to h6) and plateau state (p1 to p6) in GRS 1915+105. Following \citet{Belloni00,b32}, the letters I, K, N, O, Q, S, and T stand for 10408-01, 20402-01, 30402-01, 30703-01, 40403-01, 70702-01, and 90701-01. $\nu,$ QPO RMS, and $Q$ are the frequency, fractional RMS variability, and quality factor of the QPOs (with the fundamental QPO and its first harmonic indicated by subscripts 0 and 1, respectively). $S_{\rm 15 GHz}$ is the Ryle telescope flux density at 15.2 GHz. $\alpha$ is the best fit slope of the time lag as a function of $\log$(energy).}  
\tablenotetext{a}{For this observation, an $F$-test indicates that at $>99\%$ confidence, the lag versus $\log$(energy) is better described with a broken log-linear law than a single log-linear law. The break is in the $4-6$ keV region (see also Figure 10 of \citealt{b89}).}
\end{deluxetable*}
\begin{figure*}[t]
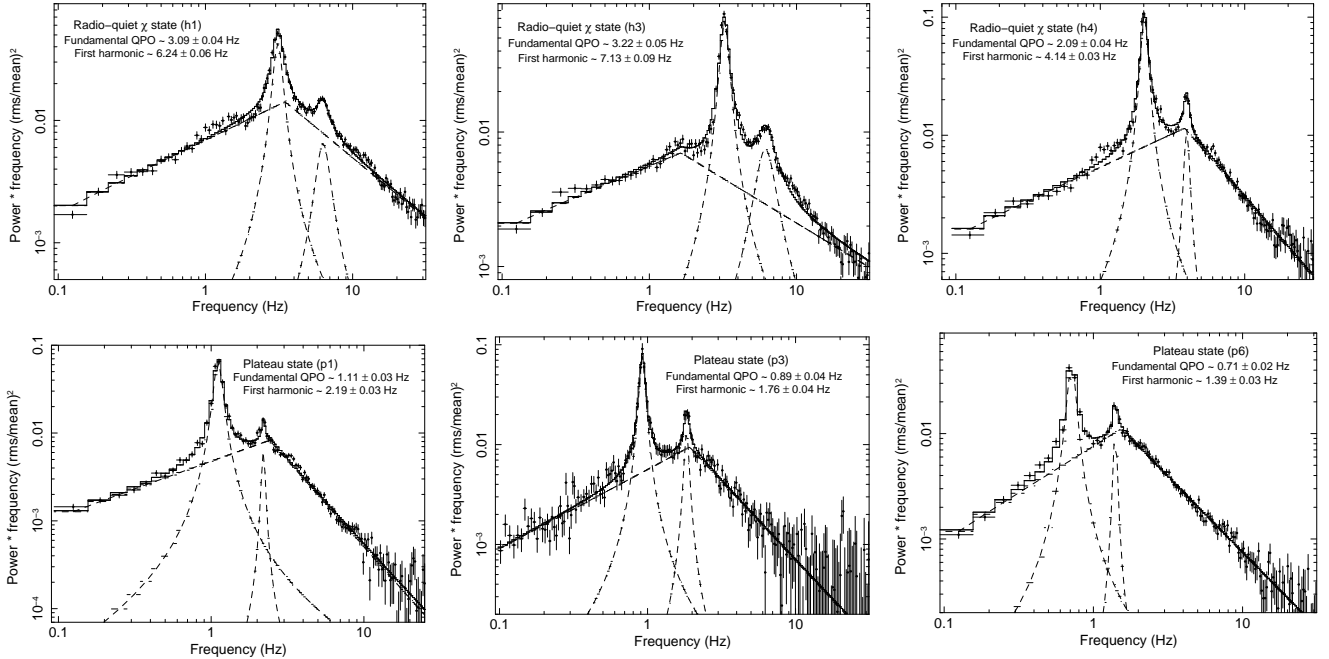

\centerline{\hfill\includegraphics[width=0.230\textwidth,angle=-90]{f3a}\hfill
\includegraphics[width=0.230\textwidth,angle=-90]{f3c}\hfill
\includegraphics[width=0.230\textwidth,angle=-90]{f3e}\hfill}\vspace{2mm}

\centerline{\hfill\includegraphics[width=0.24\textwidth,angle=-90]{f3b}\hfill
\includegraphics[width=0.234\textwidth,angle=-90]{f3d}\hfill
\includegraphics[width=0.233\textwidth,angle=-90]{f3f}\hfill}
\caption{Example RMS normalized and Poisson noise subtracted PDS in the frequency range 0.02$-$30 Hz are shown for radio-quiet $\chi$ states (left) and plateau states (right). The fitted models (dashed lines) are broken power laws plus Lorentzians for the QPOs and their harmonics.\label{fig:pds}}
\end{figure*}

\section{Observation \& data analysis}

The radio/X-ray properties of these states have been discussed by many authors, including \citet{b38} and \citet{b23}.  As noted above, plateau states are associated with strong persistent radio emission of typical 15 GHz flux density $S_{\rm 15 GHz}\sim50-100$ mJy, and are usually bracketed on either side by much brighter X-ray states and major radio flares (e.g., Figure \ref{fig:rx}). In contrast, the radio-quiet $\chi$ state is observed at both lowest observed X-ray flux (10\% $-$ 20\% of Eddington luminosity \citep{b80}) and lower radio flux from which the transition to the spectrally soft state of GRS 1915+105 is rather slow. The source remains stable in the plateau state for 5$-$20 days while it remains stable for a few months during radio-quiet $\chi$ state. The two states are simple to distinguish in the radio/X-ray correlation shown in the bottom panel of Figure \ref{fig:rx}. 

Based on these selection criteria, we choose our 12 observations from long, steady hard intervals seen by \textit{RXTE} between 1996 and 2008 that are easily identifiable as $\chi$ states (see Table 1 for details). We restrict our attention to PCU2, since it is reliably on during all observations and is best calibrated. For each observation, we extract 1 s background-subtracted light curves in the $A\equiv2-5$ keV, $B\equiv5-13$ keV, and $C\equiv13-60$ keV bands. We define a hard color as the ratio of the count rates in the $C$ and $A$ bands, and a soft color as the ratio of the count rates in the $B$ and $A$ bands. Example $2-60$ keV lightcurves, hardness-intensity diagrams, and color-color diagrams are shown in Figure \ref{fig:lchr}. In order to check whether their distinct characteristics are derived from different physical processes, we also extract X-ray spectra and power spectra. Since the X-ray spectra are generally consistent with previous work, we focus here on the timing properties.

\section{Timing Analysis} 

For our timing analysis, we extract $2.0-13.0$ keV \textit{RXTE} Proportional Counter Array (PCA) light curves from the 8 ms binned data, divide each observation into 16 s segments and calculate the power density spectrum (PDS) of each segment. As we are interested in low-frequency QPOs, we use 1/128 s time resolution so that the Nyquist frequency of each segment is 64 Hz for all observations. To attenuate noise, we compute and subtract the contribution of photon counting noise (for the low frequencies considered here, no dead time correction is necessary). Each PDS is normalized such that its integral gives the squared fractional rms variability \citep{b39}. The power spectra are then averaged together to produce one power spectrum for each observation.

We also calculate phase lags, time lags, and coherence functions (see \citealt{b40,b41,b47,b48}). We measure frequency-dependent phase/time lags between the $2.1-5.8$ keV and $6.3-8.4$ keV bands at 0.25 Hz frequency resolution.  For the energy-dependent time lags and coherences, our reference band is typically around $3-4$ keV, and we ignore bins above 19 keV due to poor counting statistics. We limit this analysis to $0.1-7$ Hz, as all our QPOs and their harmonics lie within this region, and because systematic errors and binning effects become significant below 0.1 Hz and above 10 Hz, respectively \citep{b49}. Our coherence functions are very similar for the plateau state and the radio-quiet states ($\gtrsim0.7$, decreasing slowly with energy), so we will not consider them further.

\subsection{Power Spectra}
The resulting power spectra are shown in Figure \ref{fig:pds}. In all cases a band-limited noise component (BLN) is present, together with a strong QPO and its harmonic. We find that a model consisting of a broken power law representing the BLN, plus two Lorentzian peaks for the QPOs, can fit all PDS well (the model components are also shown in Figure \ref{fig:pds}). Using fitted parameters, we calculate the fractional RMS and the quality factor of each QPO and its harmonics (see Table \ref{tbl:dist}). 
     
During both states, our power density spectra are characterized by a flat power law below the QPO frequency ($\Gamma_1$ = 0.13$-$0.32) and a steep power law above the QPO frequency ($\Gamma_2$ = 2.4$-$3.3). $\Gamma_1$ tends to increase with increasing QPO frequency, which is consistent with earlier results \citep{Reig00}. The only apparent difference between the PDS is that the QPO fundamental frequency is below 2 Hz during our plateau states and above 2 Hz during the radio-quiet $\chi$ states. \citet{Muno01} reported an anticorrelation between QPO frequency and radio flux and a strong positive correlation between QPO frequency and X-ray flux. Since our plateau states have both brighter radio emission and brighter X-ray emission, this could indicate that the radio properties are relatively more important in determining the timing properties.

\subsection{Phase Lags and Time Lags}

\begin{figure}
\centerline{\includegraphics[scale=0.38]{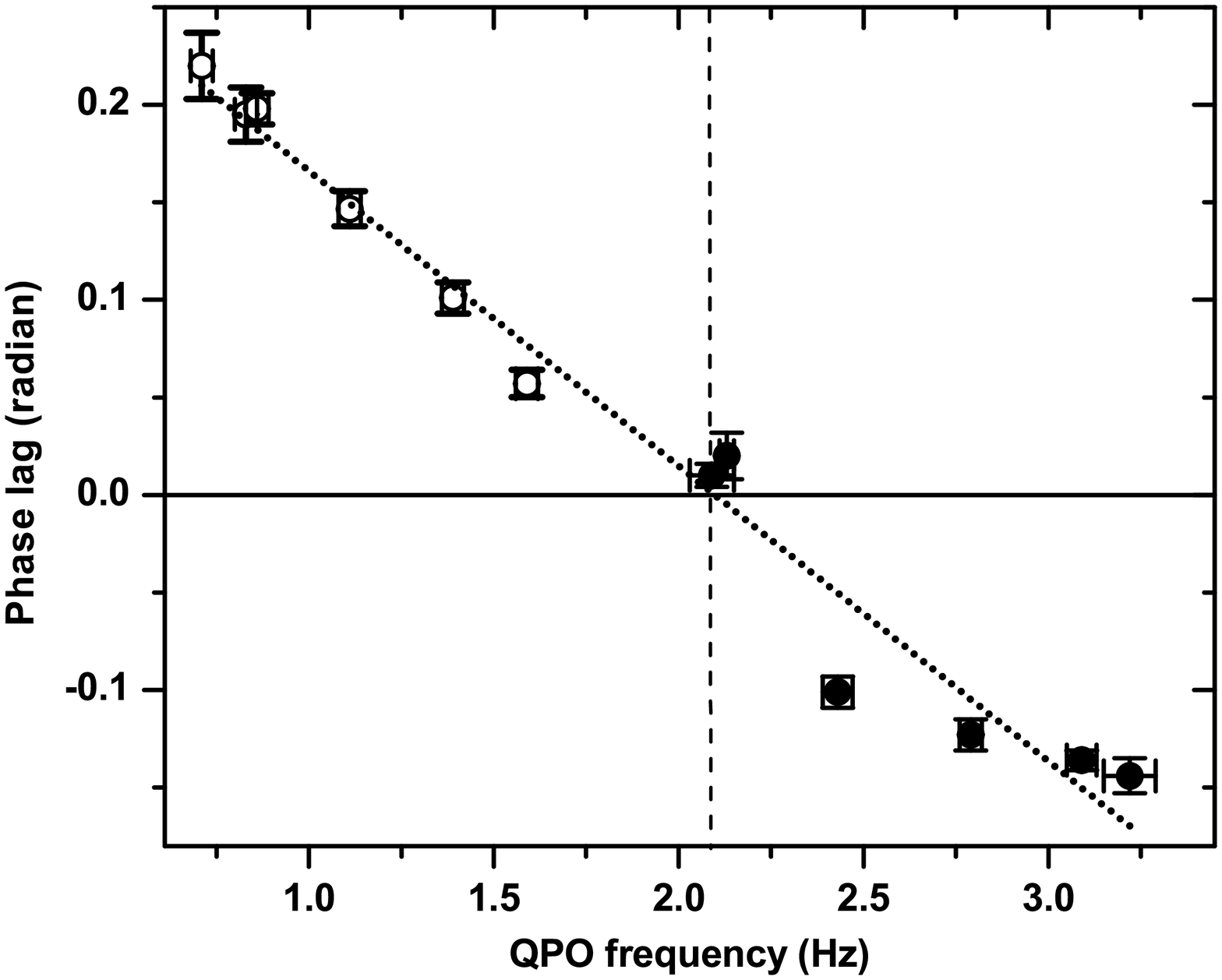}}\vspace{-10mm}
\centerline{\includegraphics[scale=0.38]{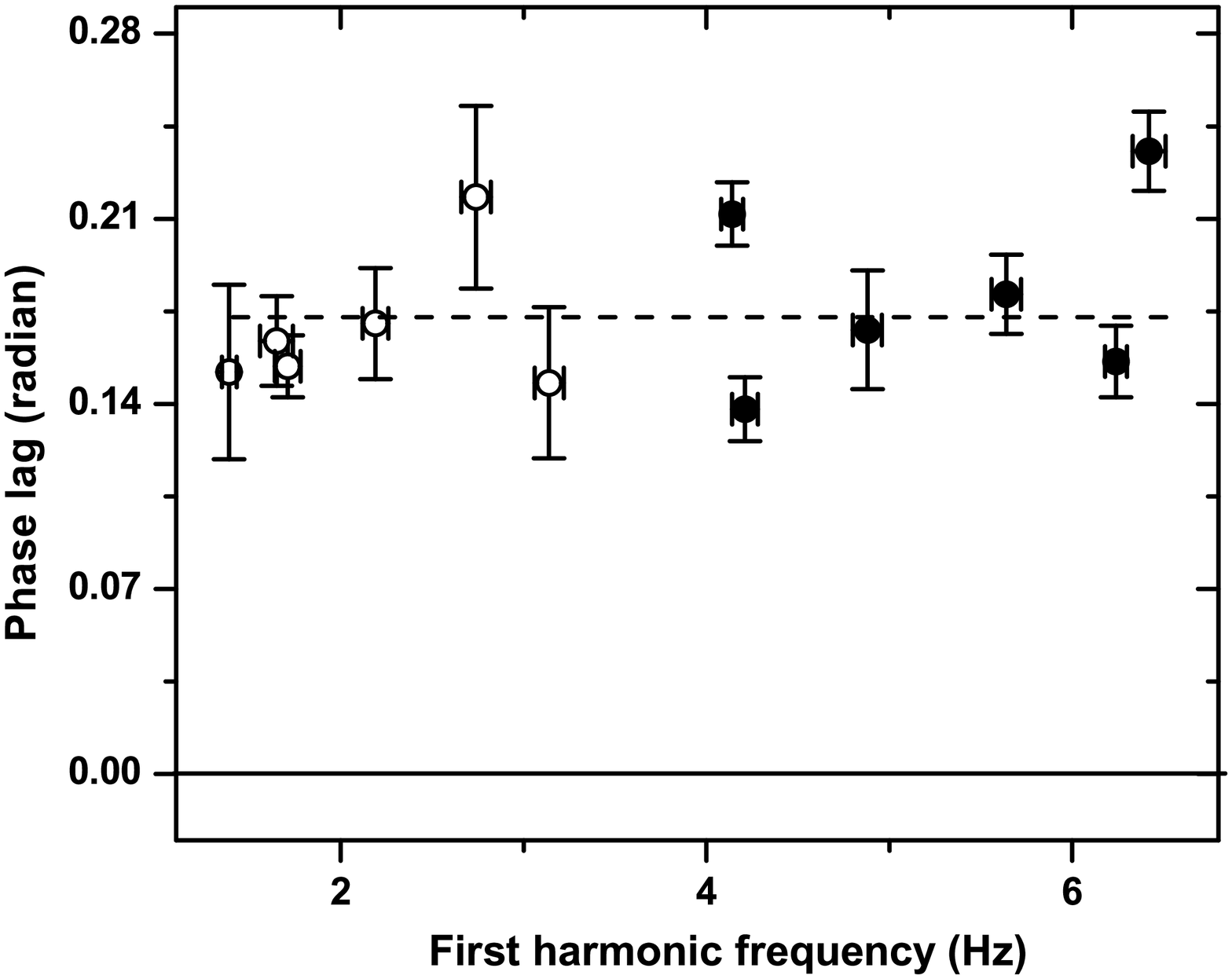}}\vspace{-3mm}
\caption{Phase lag at the QPO frequency (top) and at the frequency of the first harmonic (bottom), as a function of QPO frequency and first harmonic frequency, respectively. Lags are calculated for the 6.3$-$8.4 keV band relative to the 2.1$-$5.8 keV band. In both panels, plateau states are plotted as empty circles while radio-quiet $\chi$ states are plotted as solid circles. The phase lag at the fundamental frequency can be fitted with a straight line that intercepts the X-axis at 2.2 Hz. For the first harmonic, the lags are well-described by a constant.\label{fig:qpophase}}
\end{figure}
\begin{figure*}
\centerline{\includegraphics[width=\textwidth]{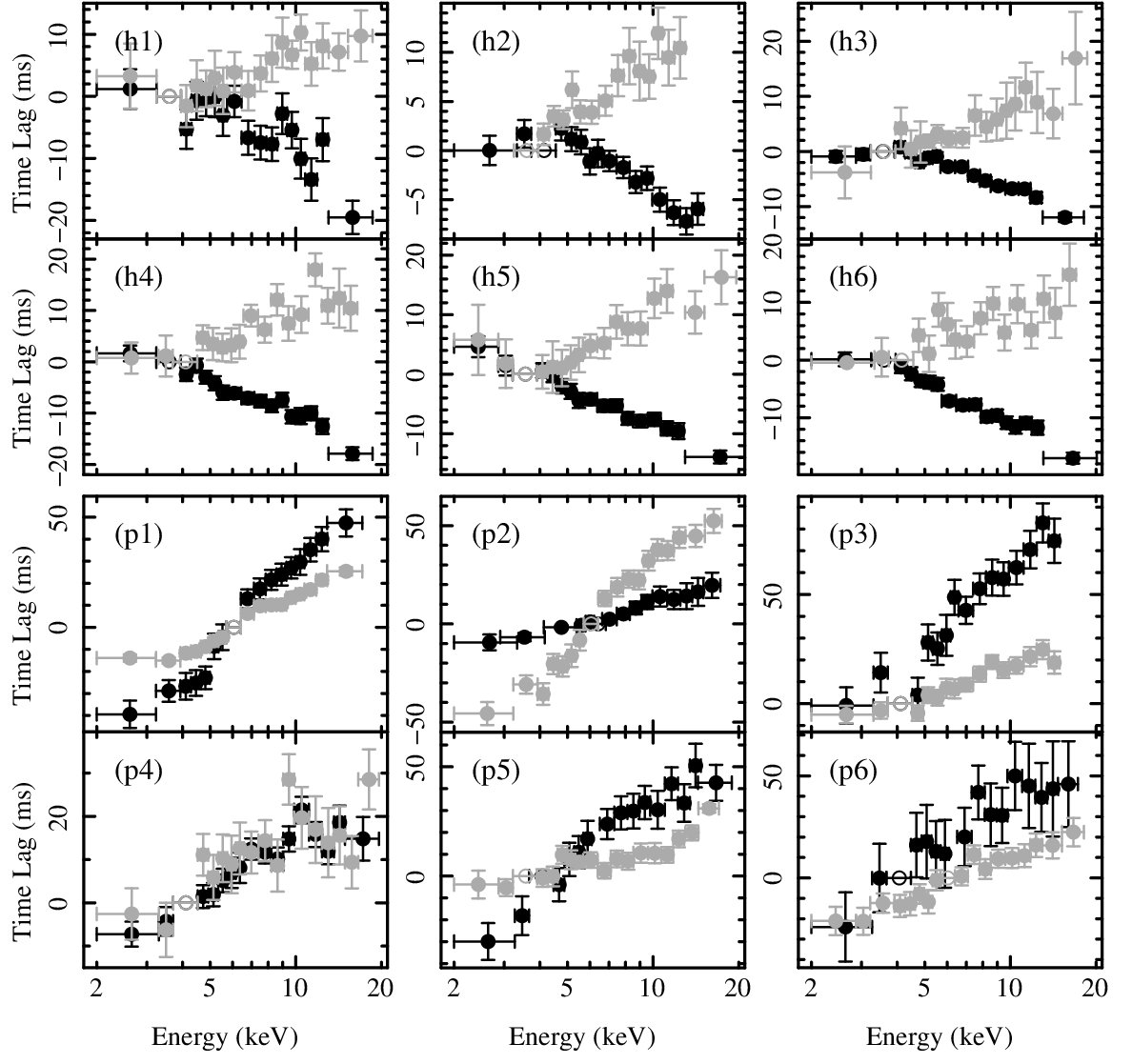}}
\caption{Energy-dependent time lags in the radio-quiet $\chi$ state (top two rows) and the plateau state (bottom two rows) of GRS 1915+105. Lags at the QPO frequency are shown in black, while lags at the frequency of the first harmonic are shown in grey. The reference bands relative to which the lags are calculated are shown as empty circles.\label{fig:tlag}}
\end{figure*}
\begin{figure*}
\centerline{\includegraphics[width=\textwidth]{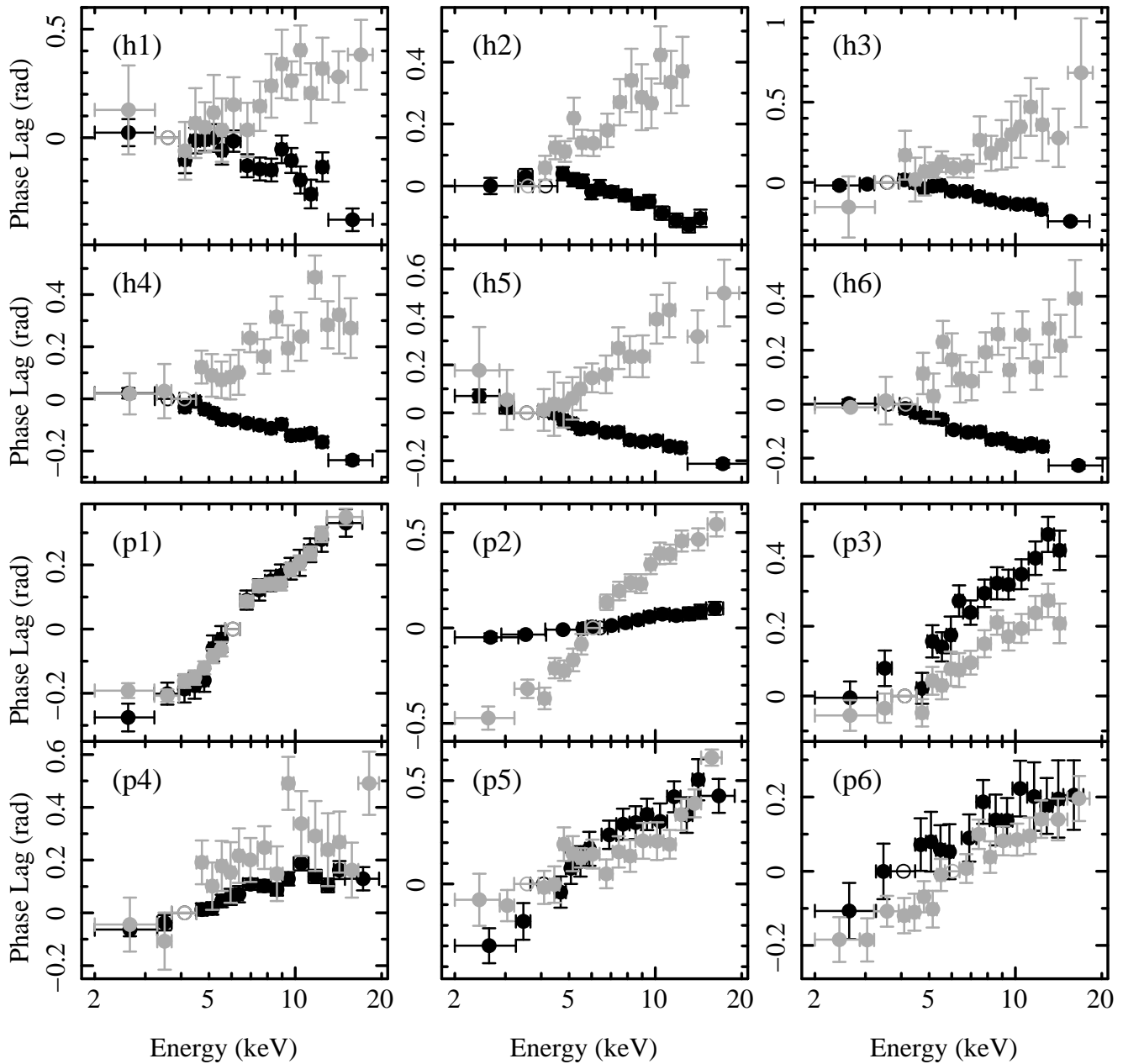}}
\caption{Energy-dependent phase lags in the radio-quiet $\chi$ state (top two rows) and the plateau state (bottom two rows) of GRS 1915+105. Lags at the QPO frequency are shown in black, while lags at the frequency of the first harmonic are shown in grey. The reference bands relative to which the lags are calculated are shown as empty circles.\label{fig:plag}}
\end{figure*}
\begin{figure*}
\centerline{\hfill\includegraphics[scale=0.35]{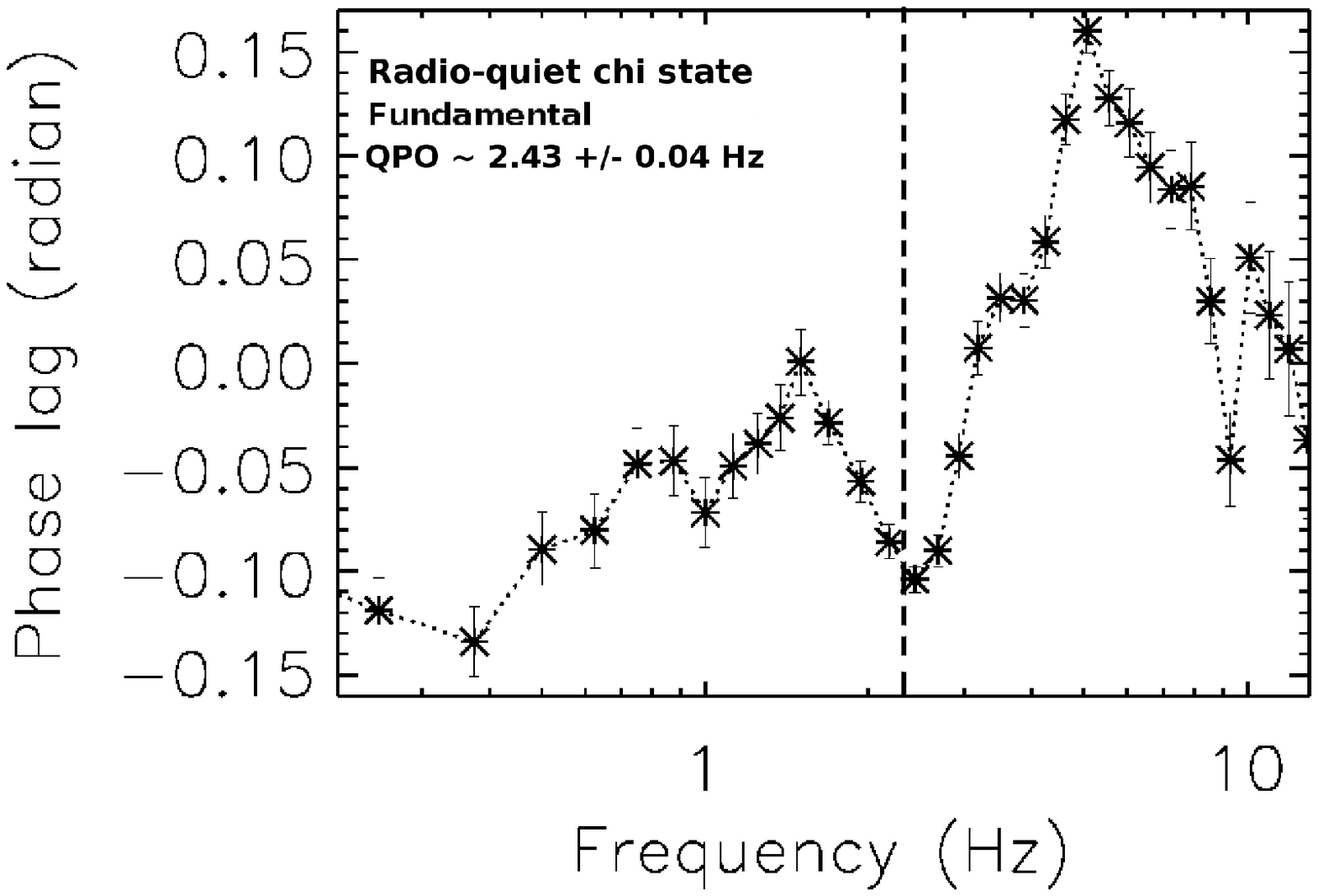}\hfill\includegraphics[scale=0.35]{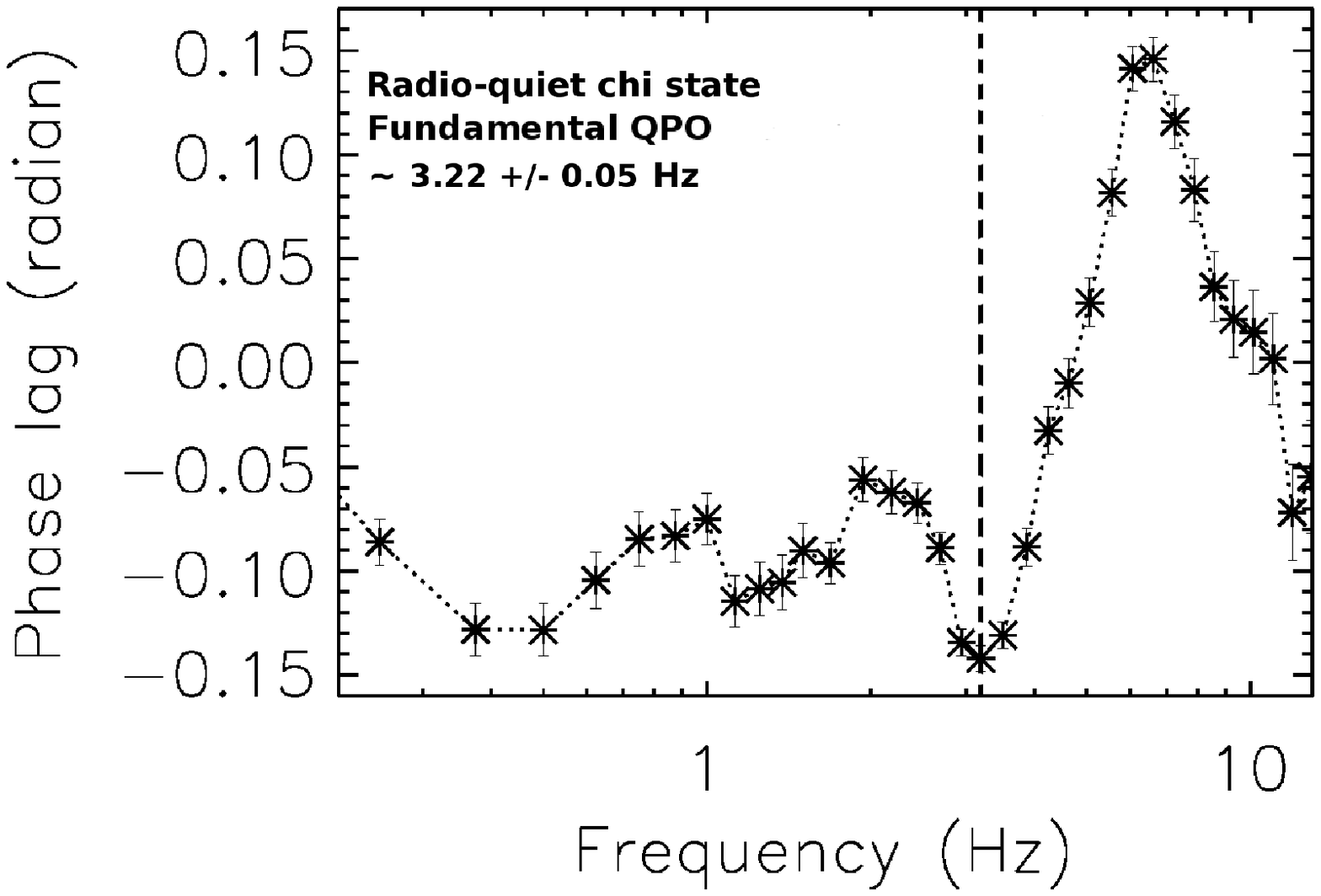}\hfill\includegraphics[scale=0.35]{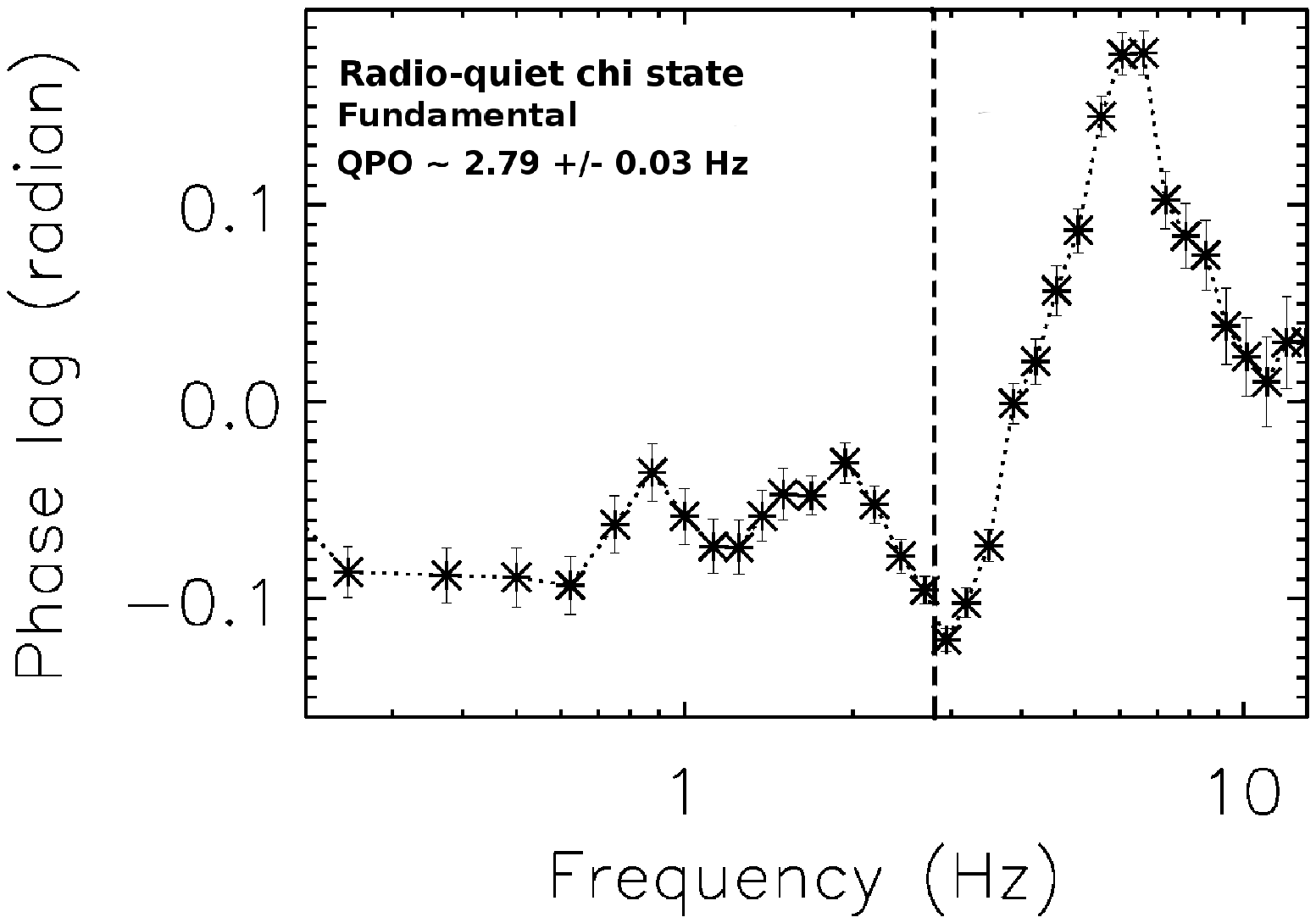}\hfill}\vspace{-1mm}
\centerline{\hfill\includegraphics[scale=0.35]{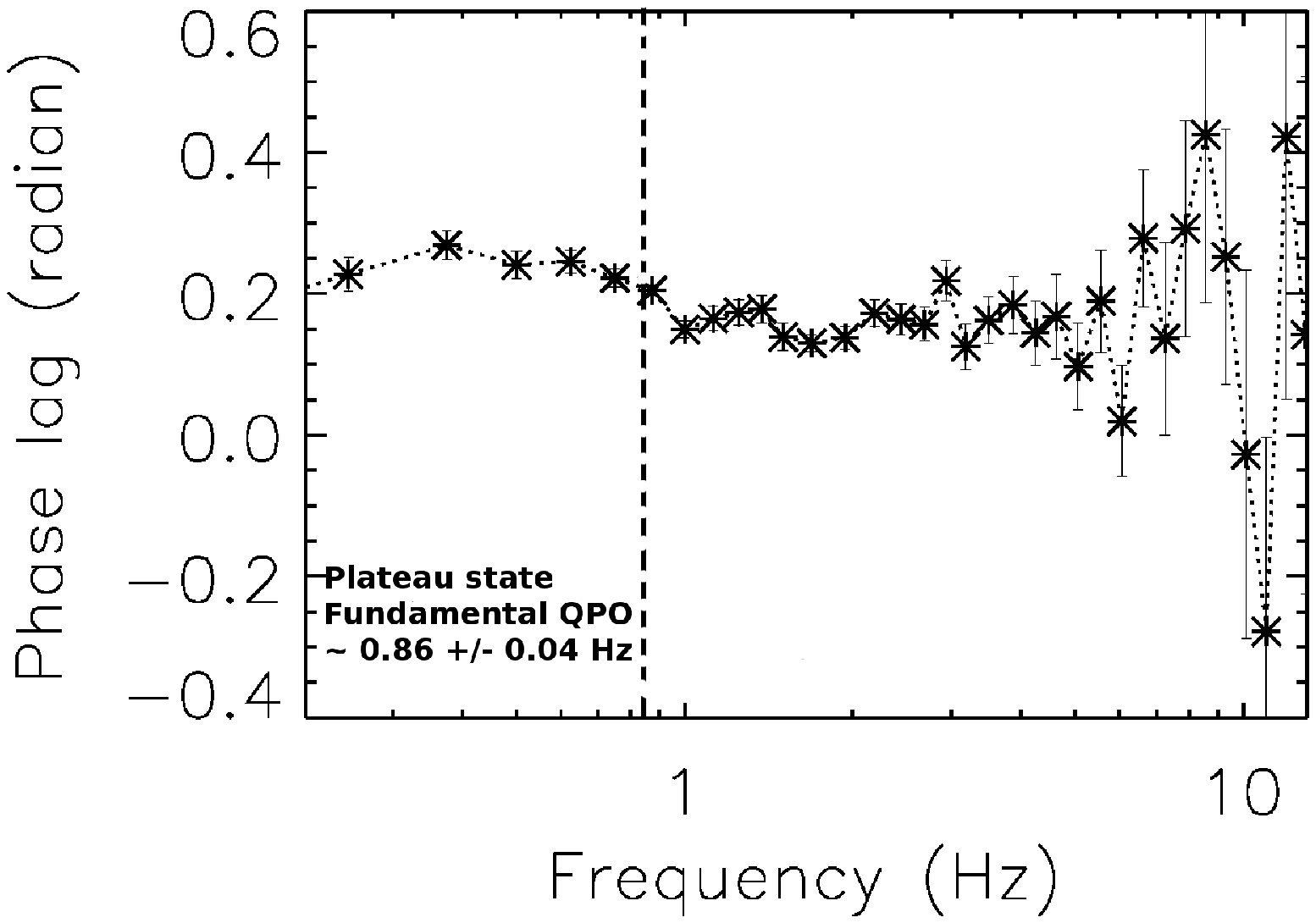}\hfill\includegraphics[scale=0.35]{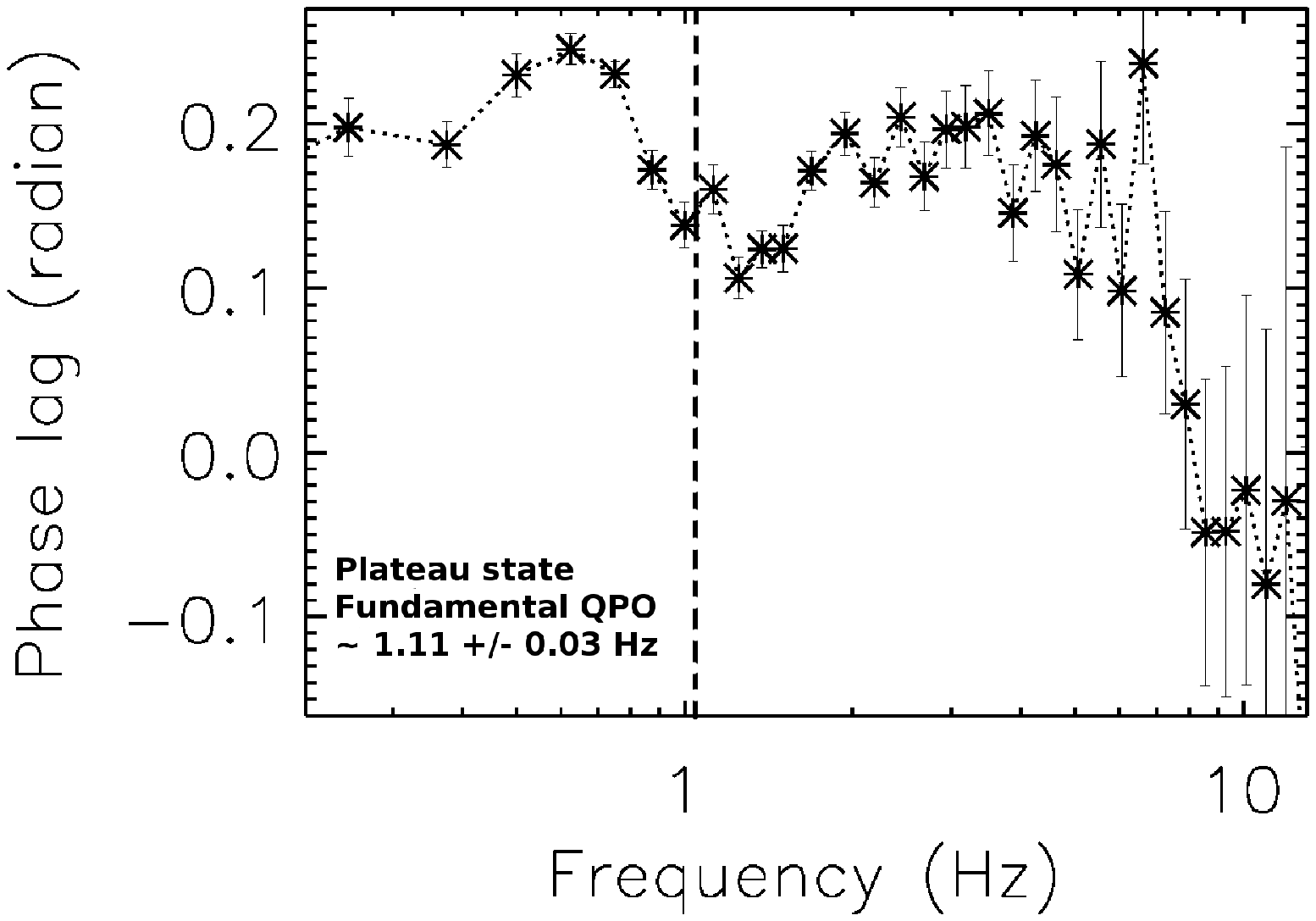}\hfill\includegraphics[scale=0.35]{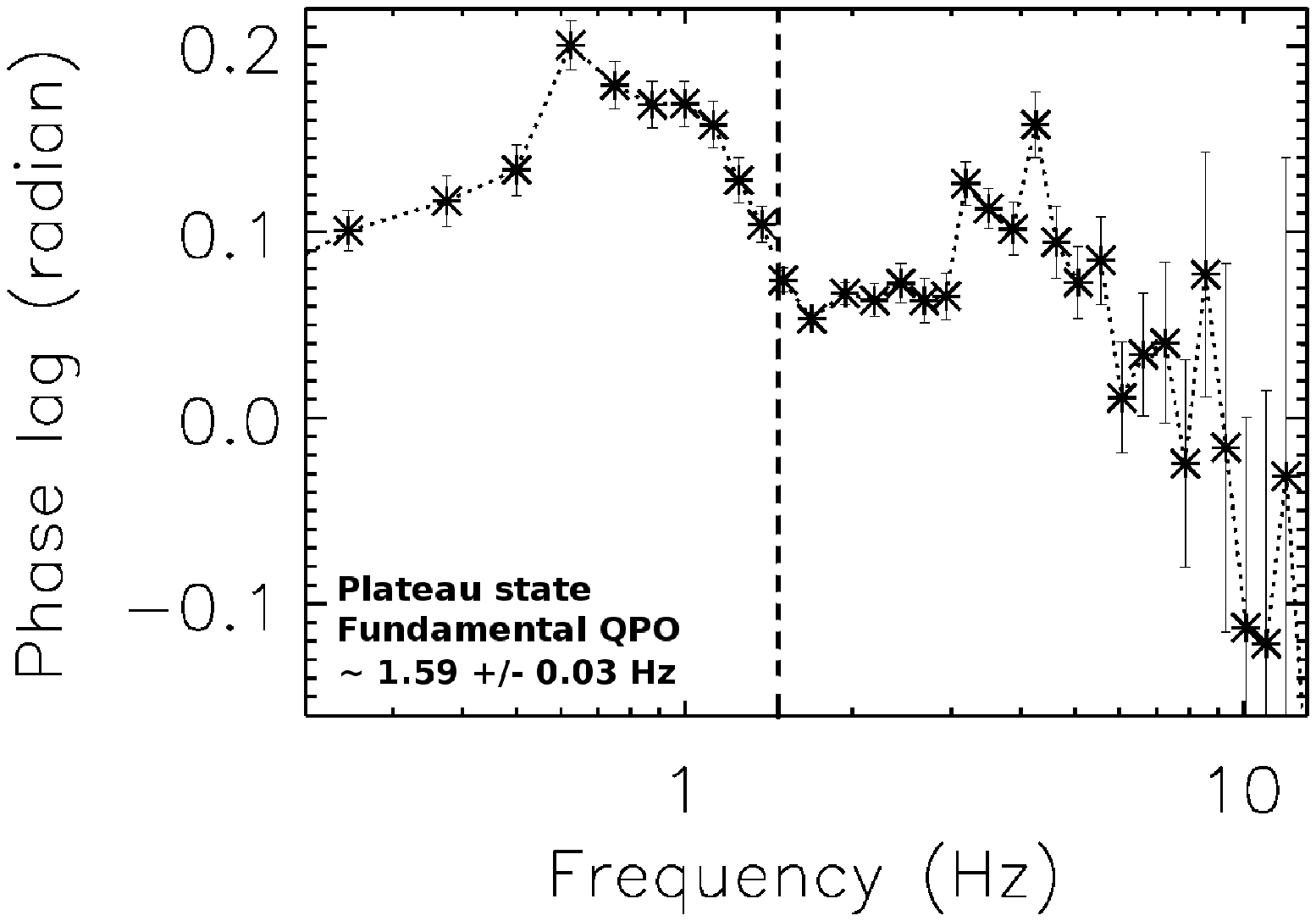}\hfill}
\caption{Time lag of the variability in the 6.3$-$8.4 keV band with respect to the variability in 2.1$-$5.8 keV band as a function of Fourier frequency during observations of GRS 1915+105 in the radio-quiet $\chi$ state (top row) and the plateau state (bottom row). In all panels, QPO frequencies are shown by dotted vertical lines.\label{fig:plagfreq}}
\end{figure*}

Following the focus of previous work (e.g.\ \citealt{Reig00}), we show the phase lag at the QPO frequency as a function of QPO frequency in the top panel of Figure \ref{fig:qpophase}. Apparent is the correlation noted above, i.e.\ a lag that decreases with QPO frequency, switching from hard to soft around a frequency of $\sim2.2$ Hz. Remarkably, however, the phase lags at the frequency of the first harmonic are (1) always positive and (2) apparently constant at $\phi_k\sim0.172$ rad. In contrast, \citet{Reig00} found that the phase lag at the frequency of the $n^{\rm th}$ harmonic changed from positive to negative around $2n$ Hz. This difference may be due to our use of different energy bands, but we shall return to this result in Section \ref{sec:discuss}.

The phase lags at the QPO frequencies demonstrate some common process between the plateau state and the radio-quiet state, but the energy-dependent lags (Figures \ref{fig:tlag} and \ref{fig:plag}) reveal significant differences. While the radio-quiet $\chi$ states show (mostly) negative time lags that decrease with energy, the plateau states show (mostly) positive lags that increase with energy. The harmonics, however, have very similar lag spectra, which is particularly interesting in light of the lag-frequency dependence in Figure \ref{fig:qpophase}. In most cases, the time lags appear to be linear in $\log$(energy). Stee Table \ref{tbl:dist} for slopes of the log-linear fits. It is notable that in two radio-quiet $\chi$ states (h2, h3) and two plateau states (p1 harmonic, p5 fundamental), the lags are better described with a $\sim4-6$ keV break at $>99\%$ confidence (see also \citealt{b89}).

We also find differences between the plateau state and radio-quiet $\chi$ state in the frequency-dependent lags (see examples in Figure \ref{fig:plagfreq}). During the plateau states, the frequency-dependent phase lags are roughly constant (especially at low frequency) at some positive value. But during the radio-quiet $\chi$ state, the lags are negative (i.e., soft) at low frequency and change to positive at a frequency that lies between the QPO and its harmonic. A dip-like feature centered on the QPO frequency is also apparent, above which the lag increases sharply. This dip-rise pattern (i.e. negative and positive lags in the QPO and harmonic, respectively) was also reported by \citet{b20}, but is not present in the plateau state.

Overall, our lag-frequency spectra are comparable to the examples given in Figure 4 of \citet{Muno01}: in the radio-quiet $\chi$ state, the lag is a generally-increasing function of frequency, but appears to be flatter during the plateau state. Our plateau state phase lag-frequency spectra are also comparable to those of \citet{Reig00}. However, our results on the energy dependence of the lags and at the QPO frequency and at the first harmonic frequency confirms the complicated relationship between the timing properties, spectral properties, and physical properties of these states. \citet{Muno01} showed that the continuum phase lags are sensitive to the X-ray flux, the radio flux, and the radio spectral index. Based on a comparison of \citet{Muno01} and \citet{Reig00} and our present results, we suggest that the shape of the lag spectra may exhibit similar dependences. Clearly a comprehensive analysis of the energy-dependent and frequency-dependent lag spectra is merited to determine precisely how the disk, the corona, and the jet contribute to the X-ray variability in the plateau states and the radio-quiet $\chi$ states.

\section{Discussion}
\label{sec:discuss}
In this paper, we have reported on some of the complex and detailed timing differences between the radio-quiet $\chi$ state and the plateau state of GRS 1915+105. We have focused on a comparison of phase lags at the QPO frequency and the energy- and frequency-dependent phase and time lags, which to date have not been completely explored. We confirm the result of \citet{Reig00} that the phase lag at the QPO frequency decreases with QPO frequency and becomes negative around 2.2 Hz. However, where \citet{Reig00} found similar results at the frequencies of the first, second, and third harmonic, we find a positive and roughly constant harmonic lag for all observations. And while our Fourier frequency-dependent phase lags are comparable to previous results in the literature (\citealt{Reig00,Muno01}), our energy-dependent lags indicate a complicated relationship between the QPOs, their lags, and the radio/X-ray properties of the black hole. For example, the lags at the QPO frequence increase and decrease with energy in the plateau states and radio-quiet $\chi$ states, respectively, but both states exhibit very similar lag-energy spectra at the frequency of the first harmonic.

The phenomenology of QPOs, lags, and accretion states in black hole X-ray binaries is enormously complex (\citealt{b35,b30,b36}). As noted above, some studies of $0.5-10$ Hz QPOs in GRS 1915+105 (e.g.\ \citealt{b20,Reig00}) have found that the lags of these QPOs change sign (from hard to soft) as the QPO frequency increases above 2 Hz; some of this phase lag may be due to variation of the QPO frequency with energy (\citealt{b31}). The harmonics also exhibit interesting behavior: \citet{b30} found that for the 67 mHz QPO, the first and third harmonics exhibited hard lags, while the fundamental and the second harmonic displayed smaller soft lags. \citet{b31} classified the QPOs into three types: (1) the 0.5$-$2 Hz QPOs, whose phase lags at the fundamental and first harmonic are both positive; (2) the 2$-$4.5 Hz QPOs, whose phase lags are negative and positive at the fundamental and first harmonic, respectively; and (3) the 4.5$-$10 Hz QPOs, which show negative phase lags at both frequencies \citep{b20,Reig00}. Similarly complex timing behavior has been observed in XTE J1550$-$564 \citep{b37}.

Part of the specific difficulty in our effort to compare the radio-quiet $\chi$ state and plateau state is related to the numerous connections between physical, spectral, and timing parameters of the accretion flow. Most of our specific conclusions are drawn from this work, \citet{Reig00}, and \citet{Muno01}. Here we outline these relationships as we see them at present:

\begin{enumerate}
\item Historically, the QPO frequency is correlated with the X-ray flux and anti-correlated with spectral hardness and radio flux; there may be an additional dependence on radio spectral index. 
Since we have discovered some X-ray faint, radio quiet observations with higher QPO frequencies than some X-ray bright, radio bright observations, the QPO frequency may actually be more sensitive to the radio flux than the X-ray flux.
\item The phase lag at the QPO frequency is tightly anti-correlated with the QPO frequency. This relation holds over a range of radio/X-ray fluxes and may be independent of the radio and spectral properties of the accretion flow.
\item The phase lag at the first harmonic is a constant, positive function of its frequency, regardless of state. The apparent differences with the anticorrelation reported by \citet{Reig00} may be due to our use of different energy bands, but a more thorough analysis of the lags at the fundamental and harmonic in both states is in order.
\item Phase lags may not be a simple function of the accretion geometry as would be determined from spectral fitting. Some complexity is required in a model-independent way by the variable lag at the QPO frequency and the constant phase lag at the first harmonic. A deeper understanding of the physical origin of QPOs and their harmonics may be necessary to resolve this difference.
\item The connection between the QPO and the harmonic is highlighted by the diversity of lag-frequency spectra (this work; \citealt{Reig00,Muno01}). Radio-quiet $\chi$ states show a sharp change in lag around the QPO, while the plateau states do not. Are these differences also related to variations in the X-ray or radio fluxes or spectral shapes? 
\item The energy-dependent lag also appears to be a function of multiple parameters. \citet{Reig00} shows that the slope of this curve depends on the frequency of the QPO. But where they show a flat lag-vs-energy curve for a 2.3 Hz QPO in the plateau state, we have a steeply declining curve for a 2.4 Hz QPO in the radio-quiet $\chi$ state, Again, it is important to measure the dependence of this timing property on the X-ray and radio behavior of the source.
\item During the plateau state, the broadband noise, QPO, and harmonic have similar lags, and the QPO and harmonic have similar lag-energy spectra. This suggests the possibility of a common radiation/variability mechanism, and should be explored further. 
\end{enumerate}

\subsection{\it Implication of hard lag and soft lag in the light of some generic models}

Several efforts have been made to explain hard and soft lag observed in GRS 1915+105 and canonical low hard state of a few more BHXBs. For interpreting the hard lag, `sphere + disk' Comptonization model \citep{b41} is used frequently which assume the Comptonization of soft, seed photons from the inner disk by a spherical, hot electron plasma, called `corona'. Although this model qualitatively explain the shape of phase lag spectra, in terms of light travel time, the observed lag time-scale of radio-quiet $\chi$ state and plateau state (up to $\sim$100 ms) from our analysis requires the corona size up to thousand gravitational radii which is non-physical. Similarly, reflection of Comptonized photons from the disk or reverberation lags has been invoked to explain both hard and soft lags for Active Galactic Nuclei \citep{b53, b69}. Depending upon plasma optical depth and temperature gradient in Comptonizing region, \citet{Reig00} proposed two different mechanisms of Compton scattering which leads to Comptonization delays that may also give rise to hard and soft lags. soft lag due to Comptonization delays is also explained by a Compton up-scattering model \citep{b93}, where oscillations in plasma temperature is responded by the variation in the Wien blackbody temperature of the soft seed photons. These lags are due to light travel time effects and are of the order of the light crossing time of the system. Thus simple reflection or Comptonization delays cannot account for observed lag time-scales. 

Lags that are much longer than the expected light-crossing times are also seen in the hard states of BHXBs (e.g. \citealt{b41, b68, b66}), where in the 3-20 keV range, harder energies always lag softer energies, unlike the lags seen here in the GRS~1915+105 QPOs.  The currently favored interpretation of these canonical hard-state lags are that they are associated with inward propagation of mass accretion fluctuations through the accretion flow, where the harder emission is assumed to originate at smaller radii \citep{b89, b67}. Recently, \citet{b66} found that at frequencies below $\sim$1 Hz there is a sharp downturn in the lag-energy spectrum below $\sim$ 2 keV, where the accretion disk blackbody emission appears in the time-averaged spectrum, i.e., the variations in disk photons significantly lead those of the power law photons by a few tenths of a second.  This result strongly supports the propagating fluctuations idea, implying that the lags are produced by fluctuations in the accretion flow that start out in the blackbody emitting disk.  At frequencies above 1~Hz, the lag behavior switches around, such that the disk photons start to lag the harder, power law photons, with lags of a few ms (which \citealt{b66} interpret in terms of the X-ray heating of the disk by the power law, with lags now comparable to the light-crossing time between the power law emission component and the disk).

It is interesting to consider whether the same model could explain the QPO lags observed here, where we also find evidence for a switch in the lag behavior (above $\sim$ 2.2 Hz) and most notably, a possible break in the lag-energy dependence at lower energies (below $\sim4-6$ keV of at least one observation, which may correspond to the hotter disk expected in these higher-luminosity states).  Although no detailed models have been proposed for QPOs generated by a fluctuating accretion flow, it is natural to suppose that a QPO would be produced on a specific time-scale if an instability occurs at a particular disk radius or narrow range of radii.  However, although this picture is superficially attractive, we do not favor it here because the QPO behavior in GRS 1915+105 shows a distinct difference from that seen in hard state BHXRBs, in that at high QPO frequencies the sign of the lag-energy dependence changes from hard to soft when measured between all energies that we can probe.  In contrast, \citet{b66} found that only the sign of the lags in the softest energies changed relative to the energies above 2~keV, but when measured between a pair of energies above $\sim$ 2 keV, the hard lags persisted. In the canonical hard states the change in lag-energy corresponds to a `pivoting' of the downturn to an upturn around an energy of $\sim$ 2 keV, whereas we observe something more like a `reflection' of the lag-energy spectrum seen at low frequencies.

This `reflection' type evolution of the lag-energy spectrum is significant, because it suggests that, rather than representing a distinct change in the underlying physical mechanism causing the QPO lags, the change at $\sim$ 2.2 Hz is more like a simple `phase rotation' of the same underlying mechanism.  We do not yet have a detailed physical model that can explain this lag-energy behavior, but we note that it is highly suggestive that any viable model must involve some sort of rotational effect, i.e., a rotation of the emitting region structure leading to a change in the phase offset of soft and hard photons, rather than a change in hard vs. soft emissivity profile to explain the lags and their evolution with QPO frequency.  More detailed study is required, but we note that one model which could potentially satisfy this requirement is the Lense-Thirring precession model for the QPOs \citep{b45}, where the QPO is interpreted in terms of precession of the inner hot flow around the spin axis of the black hole.

\section{Acknowledgements}

We are thankful to the anonymous referee for his/her constructive comments. This research has made use of the General High-energy Aperiodic Timing Software (GHATS) package developed by T.M. Belloni at INAF - Osservatorio Astronomico di Brera and the data obtained through the High Energy Astrophysics Science Archive Research Center online service, provided by the NASA/Goddard Space Flight Center. J.N. gratefully acknowledges funding support from NASA through the Einstein Postdoctoral Fellowship, grant PF2-130097, awarded by the Chandra X-ray Center, which is operated by the Smithsonian Astrophysical Observatory for NASA under contract NAS8-03060, and from NASA through the Smithsonian Astrophysical Observatory contract SV3-73016 to MIT for support of the Chandra X-ray Center, which is operated by the Smithsonian Astrophysical Observatory for and on behalf of NASA under contract NAS8-03060.

\clearpage 

\end{document}